\documentclass[final]{svjour2}
\usepackage{graphicx}
\usepackage{rotating}
\usepackage{amssymb}
\usepackage{mathptmx}
\usepackage[numbers]{natbib}

\usepackage{graphicx}
\usepackage{dcolumn}
\usepackage{bm}
\usepackage{color}

\makeatletter
\journalname{Journal of Low Temperature Physics}

\bibpunct{}{}{,}{s}{}{,}

\begin{document}

\newcommand{\hdblarrow}{H\makebox[0.9ex][l]{$\downdownarrows$}-}
\title{Tachyon Condensation and Brane Annihilation in Bose-Einstein Condensates:
Spontaneous Symmetry Breaking in Restricted Lower-dimensional Subspace}

\author{Hiromitsu Takeuchi$^1$ \and Kenichi Kasamatsu$^2$ \and Makoto Tsubota$^{3,4}$ \and Muneto Nitta$^5$}

\institute{1:Graduate School of Integrated Arts and Sciences, Hiroshima University, Japan\\
\email{hiromitu@hiroshima-u.ac.jp}\\
2: Department of Physics, Kinki University, Japan\\
3: Department of Physics, Osaka City University, Japan\\
4: The Osaka City University Advanced Research Institute for Natural Science and Technology (OCARINA), Osaka City University, Japan\\
5: Department of Physics and Research and Education Center for Natural Sciences, Keio University, Japan}

\date{\today}

\maketitle


\begin{abstract}
In brane cosmology, the Big Bang is hypothesized to occur by the annihilation of the brane--anti-brane pair in a collision,
 where the branes are three-dimensional objects in a higher-dimensional Universe.
 Spontaneous symmetry breaking accompanied by the formation of lower-dimensional topological defects, {\it e.g}. cosmic strings, is triggered by the so-called `tachyon condensation', where the existence of tachyons is attributable to the instability of the brane--anti-brane system.
 Here, we discuss the closest analogue of the tachyon condensation in atomic Bose--Einstein condensates.
 We consider annihilation of domain walls, namely branes, in strongly segregated two-component condensates,
where one component  is sandwiched by 
two domains of the other component.
In this system, the process of the brane annihilation can be {\it projected} effectively as ferromagnetic ordering dynamics {\it onto} a two-dimensional space.
Based on this correspondence, three-dimensional formation of vortices from a domain-wall annihilation is considered to be a kink formation due to spontaneous symmetry breaking in the two-dimensional space.
We also discuss a mechanism to create a `vorton' when the sandwiched component has a vortex string bridged between the branes.  
We hope that this study motivates experimental researches to realize this exotic phenomenon of spontaneous symmetry breaking in superfluid systems.

\end{abstract}

\section{Introduction}
While a tachyon is a hypothetical superluminal particle in special relativity,
a tachyon field can exist 
in quantum field theories,
as a consequence of the instability of the system.
Here, the state considered is initially at a local maximum of an effective potential $V(T)$,
and then the tachyon `rolls down' toward a minimum of the potential,
 which causes the exponential amplification of a tachyon field $T$ from the initial state $T=0$.
The rolling-down process is called tachyon condensation.
In string theory, which is the most promising 
candidate of the unified theory, 
tachyon condensation occurs 
in a system containing a Dirichlet(D-)brane and an anti-D-brane 
\cite{Sen:2005}.
Here, the D-brane is solitonic excitation \cite{Polchinski:1998}
and the anti-D-brane is its anti-object
which annihilates the D-brane in collision.
The system falls into the true vacuum when the annihilation is completed.

Tachyon condensation of the brane annihilation is applied to brane cosmology \cite{Dvali:2001,Langlois:2002, Quevedo:2002, McAllister:2008}
in which phase transitions in the early Universe
are supposed to occur as a result of a collision of branes.
Tachyon condensation causes formations of lower-dimensional branes after the collision,
which corresponds to nucleation of cosmic strings due to the phase transitions accompanied by spontaneous symmetry breaking (SSB) in brane cosmology \cite{Jones:2002,Sarangi:2002,Dvali:2004}.
This phenomenon resembles the Kibble-Zurek mechanism \cite{Kibble:1976, Zurek:1985}, which explains formations of topological defects due to conventional SSB in the early Universe \cite{Vilenkin:1994}.
However, SSB due to a brane annihilation is different from the conventional SSB in the sense that phase ordering occurs in a lower-dimensional subspace embedded in higher dimensions.
The understanding of brane annihilation in terms of a phase transition has not been yet clarified
 since how the presence of the extra-dimensions influences the phase ordering dynamics is not clear, 
{\it e.g.} the distance or relative velocity between the two branes 
and deformations of the branes into the extra-dimensions.
Although the Kibble-Zurek mechanism has been tested thoroughly in condensed matter laboratory experiments 
\cite{Hendry:1994,Bowick:1994,Bauerle:1996, Ruutu:1996,Carmi:2000,Maniv:2003,Monaco:2006,Sadler:2006,Weiler:2008},
brane annihilations as {\it subspatial} SSB phenomena are quite a novel concept in condensed matter systems.

\begin{figure} [tpbh] \centering
  \includegraphics[width=.7 \linewidth]{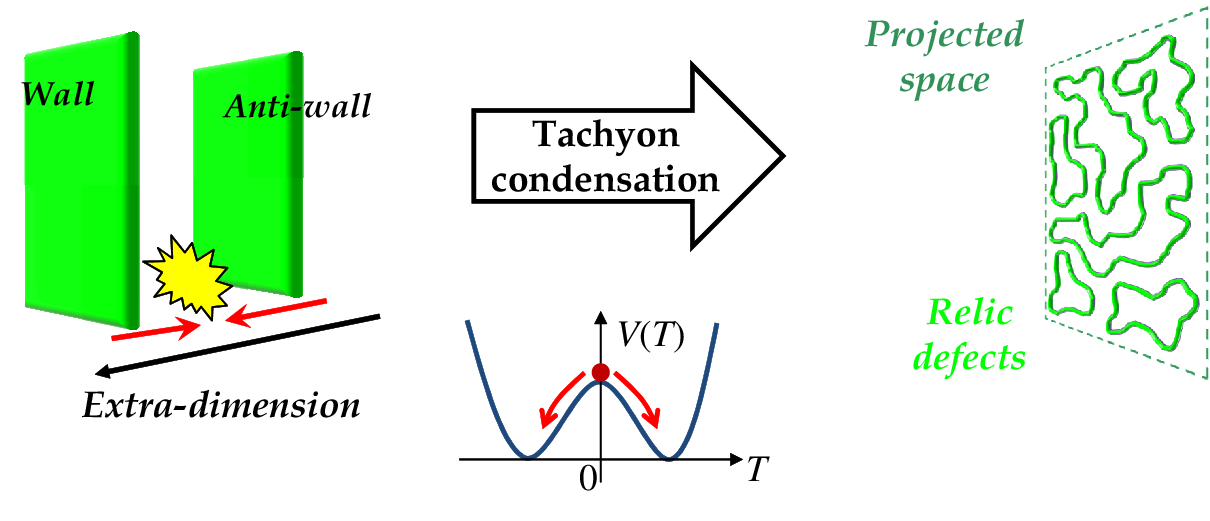}
  \caption{Schematic diagrams of a domain wall annihilation.
}
\label{fig:Schematic}
\end{figure}

Recently, we showed that vortex formations via the pair annihilation of domain walls in two-component Bose-Einstein condensates (BECs) in three-dimensions are regarded as kink formations in the effective tachyon field defined in the projected-2D space \cite{Takeuchi:2012}, as shown schematically in Fig.~\ref{fig:Schematic}.
 To the best of our knowledge, this system is the first example of {\it subspatial} SSB phenomena in condensed matter system.
 In this system, we can address the nonlinear dynamics of the brane annihilation, such as defect nucleation and the subsequent effective phase ordering dynamics,
 which is difficult in string theory and brane cosmology.
 In this paper, we briefly review a series of our works on annihilations of domain walls, a `branes' and `anti-brane', in two-component BECs  \cite{Takeuchi:2012, Takeuchi:2011, Nitta:2012}.

This paper is organized as follows.
 In Sec.~\ref{sec:2}, we introduce the Gross-Pitaevskii model and consider a state with a domain wall and an anti-domain wall in strongly segregated two-component BEC,
 where a component is sandwiched between two domains of the other component.
 In Sec.~\ref{sec:3}, vortex formation from an domain wall annihilation is discussed.
 In Sec.~\ref{sec:4}, the annihilation process is reviewed by introducing tachyon field in the projected-2D conceptually.
 In Sec.~\ref{sec:5}, we discuss the influence of motion of the sandwiched component along the nucleated vortices.
Section~\ref{sec:6} is devoted to a summary and discussion.

\section{Domain wall and anti-domain wall in two-component BECs}\label{sec:2}
We consider two-component BECs, which consist of condensations of two distinguishable Bose particles.
Two-component BECs are well described by the macroscopic wave function as two complex order parameters, $\Psi_1$ and $\Psi_2$, in the Gross-Pitaevskii model at zero temperature \cite{Pethick:2002}.
 The wave functions obey the action
${\cal S}=\int dt \int d^3x\left[ i\hbar\left(\Psi_1^*\partial_t \Psi_1+\Psi_2^*\partial_t \Psi_2\right)-{\cal K}-{\cal V} \right]$
with kinetic energy density ${\cal K}$ and potential energy density ${\cal V}$;
\begin{eqnarray}
{\cal K}&=&\frac{\hbar^2}{2m_1}|{\bf \nabla}\Psi_1|^2+\frac{\hbar^2}{2m_2}|{\bf \nabla}\Psi_2|^2, 
\label{eq:V_kin}\\
{\cal V}&=&\frac{1}{2}g_{11}|\Psi_1|^4+\frac{1}{2}g_{22}|\Psi_2|^4 +g_{12}|\Psi_1|^2|\Psi_2|^2-\mu_1 |\Psi_1|^2-\mu_2 |\Psi_2|^2,
\label{eq:V_bulk}
\end{eqnarray}
where we used the coupling constant $g_{jk}=2\pi\hbar^2 a_{jk}/m_{jk}~(j,k=1,2)$ with the reduced mass $m_{jk}^{-1}=m_j^{-1}+m_k^{-1}$ and the $s$-wave scattering length $a_{jk}>0$ between atoms in the $\Psi_j$ and $\Psi_k$-components. The chemical potential $\mu_j>0$ is introduced as the Lagrange multiplier for the conservation of the norm $N_j=\int  d^3x |\Psi_j|^2$.
Two-component BECs undergo phase separation for $g_{12}>\sqrt{g_{11}g_{22}}$ making domain structures with $\Psi_j$ domains ($j=1,2$), in which the $\Psi_j$-component is dominant and the $\Psi_{k\neq j}$ component vanishes. 
 We suppose, for example, the following parameters, $m_1=m_2\equiv m$, $g_{11}=g_{22}\equiv g$, $ \mu_2/\mu_1\equiv\nu$, and $ g_{12}/g\equiv\gamma=2$;
this is experimentally feasible, {\it e.g.} Ref.~\cite{Papp:2008}.
We consider the domain structure, where a $\Psi_2$ domain is sandwiched by two $\Psi_1$ domains.
The length scale of this system is characterized by the healing length $\xi \equiv \hbar/\sqrt{m \mu_1}$ of the $\Psi_1$ component.

As a first step of the analysis, we consider  the energy density in bulk [the potential ${\cal V} (\Psi_1,\Psi_2)$ of (\ref{eq:V_bulk})] 
by neglecting the contribution from the kinetic energy (\ref{eq:V_kin}).
The amplitude $|\Psi_2|$ is determined to minimize ${\cal V}$ with $\Psi_1$ fixed.
Then the density $n_2(\Psi_1)=|\Psi_2|^2$ is parameterized by $\Psi_1$ as $n_2=(\mu_2-g_{12}|\Psi_1|^2)/g$ for $\mu_2>g_{12}|\Psi_1|^2$ and $|\Psi_2|^2=0$ for $\mu_2<g_{12}|\Psi_1|^2$.
 By inserting this into ${\cal V}$,
 the potential ${\cal V}$ is reduced to a function $W$ of $|\Psi_1|$, $W(|\Psi_1|) \equiv {\cal V} (\Psi_1,\sqrt{n_2(\Psi_1)})=w_4|\Psi_1|^4+w_2|\Psi_2|^2+w_0$.
 We do not consider the case with $\nu < 1/\gamma$
 since then the potential $W$ with $w_2<0$ has a local maximum at $\Psi_1=0$ and
 $\Psi_2$ domains cannot exist in bulk.
 Since the coefficient $w_4$ is negative for $\mu_2>g_{12}|\Psi_1|^2$ but positive for $\mu_2<g_{12}|\Psi_1|^2$,
the potential $W$ has a local (global) minimum at  $|\Psi_1|=\sqrt{\mu_1/g}$ for $\nu>1$ ($\nu<1$).
On the other hand, the state with $|\Psi_2|^2=\mu_2/g$ and $\Psi_1=0$, realized in the $\Psi_2$ domain, is a local (global) minimum of the bulk potential for $\nu<1$ ($\nu>1$).

 Let us consider a pair of flat domain walls, our brane and anti-brane, in an uniform system as is illustrated in Fig.~\ref{fig:VortexFormation} (top left).
 If the thickness of the domain walls is neglected,
 a stable domain structure is determined from the balance between the surface pressures and the interface tension of the walls.
  The (hydrostatic) pressure is defined as $P_j=g_{jj}n_j^2/2=\mu_j^2/2g_{jj}$ in $\Psi_j$ domain.
 We impose the boundary condition $|\Psi_1(z\to \pm \infty)|=\mu_1/g_{11}$,
 which can be regarded as an external pressure $P_1=\mu_j^2/2g_{jj}$ in this system.
 A flat domain wall is stabilized for $P_1=P_2$ with $\nu=1$.
 However, for $\nu<1$, the $\Psi_2$ domain between the two $\Psi_1$ domains can be crushed due to the pressure imbalance, $P_1>P_2$, which leads to annihilation of the domain walls.
 This means that, for $\nu<1$,
 the $\Psi_2$ domain is destabilized due to the existence of domain walls although the domain itself is locally stable far from the walls.

The annihilation should be related to the interaction between the branes. 
 Since the `penetration' of the amplitude $|\Psi_1|$ ($|\Psi_2|$) decays exponentially from the domain wall into the $\Psi_2$ ($\Psi_1$) domain,
 the short-range interaction between the branes is effective only when the inter-brane distance $R$ is comparable to the `penetration depth', called the brane thickness.
Here, we define the inter-brane distance $R$ as the distance between the two $|\Psi_1|=|\Psi_2|$ planes at $z=\pm R/2$.
 In our case of strong segregated two-component BECs, the brane thickness is of order of healing length $\sim \xi$.
 Here, we focus on drastic annihilation processes caused by the strong inter-brane interaction,
 where the brane and anti-brane are brought close to each other at a small $R$ at $t=0$.
 Such a situation can be realized by preparing a stable pair of branes with sufficiently large $R$ and 
reducing rapidly the population of the sandwiched component.

  \section{Vortex formation from domain wall annihilation}\label{sec:3}
 
\begin{figure} [tpbh] \centering
  \includegraphics[width=.8 \linewidth]{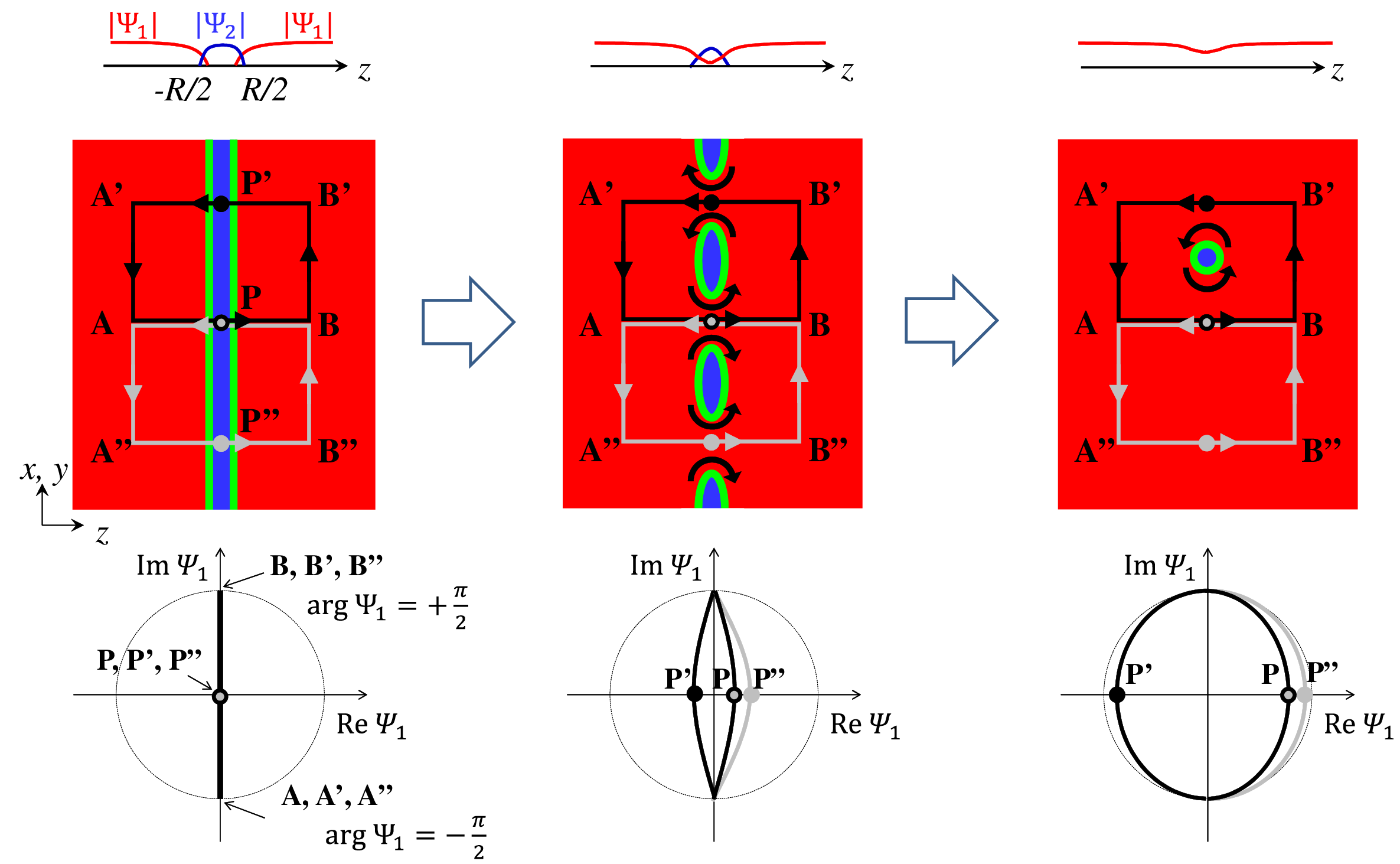}
  \caption{ Schematic diagram of vortex formations from domain wall annihilation.
(top) 1D diagrams of domain wall annihilation.
(middle) 2D diagrams of vortex formations from domain wall annihilation.
(bottom) Vortex formation in the 2D diagrams are explained in the configuration space of $\Psi_1$.
  In the initial state, the positions P, P', and P'' in the 2D diagram (middle left) is located at the origin in the configuration space (bottom left).
 The direction of motions of the points in the configuration space (bottom center) is determined by the directions of superfluid current though each junction (middle center).
 The circulation along the loop A-B-B''-A''-A has no winding number, but the loop for A-B-B'-A'-A has contains a quantized vortex (bottom right), whose core is filled with the $\Psi_2$ component (middle right) . 
}
\label{fig:VortexFormation}
\end{figure}

 The conventional process of the pair annihilation is that
 the brane and anti-brane collide straightforwardly to leave only the $\Psi_1$ component as shown in the 1D diagram of Fig.~\ref{fig:VortexFormation} (top).
 Naively, this process is forbidden because the norm $N_2$ is conserved during the process.
 However, the annihilation process can occur locally by pushing away the $\Psi_2$ component.

In general, the brane annihilation may occur inhomogeneously in various places between the branes.
Then, the annihilation processes become nontrivial by considering the phase difference between the two $\Psi_1$ domains.
 We introduce the phase difference $\Delta\Theta$ with $\arg\Psi_1(z = \pm \infty)=\pm\Delta\Theta/2$.
 The annihilation for $\Delta\Theta=\pi$ is just a case, which was realized experimentally by Anderson {\it et al.} \cite{Anderson:2001}.
In the experiment \cite{Anderson:2001},
they created a dark soliton in one component, where the phase of the order parameter jumps by $\pi$ discontinuously across the soliton like a nodal plane, and the density dip created by the soliton was filled with the other component.
 Then the filling component was selectively removed, which corresponds to the coincident limit $R\to 0$ without the $\Psi_2$ component in our model.

The inhomogeneous annihilation is initiated by making small junctions, which connect locally the two $\Psi_1$ domains by pushing out the $\Psi_2$ component in the region $-R/2 \lesssim z \lesssim R/2$.
 Since the phase $\arg\Psi_1$ must be continuously connected from a domain to the other domain through a junction,
 the local annihilation causes a superfluid current along the $z$-axis with a current velocity
 \begin{eqnarray}
v_z  \sim \frac{\hbar}{m}\frac{\Delta\Theta}{R}>0
\ \ \ \ \ {\rm or}\ \ \ \ \ 
v_z  \sim \frac{\hbar}{m}\frac{\Delta\Theta-2\pi}{R}<0
\label{eq:v_z}
\end{eqnarray}
 [see Fig.~\ref{fig:VortexFormation} (middle and bottom)].
 When the current velocities have the same sign between two neighboring junctions (P and P" in Fig.~\ref{fig:VortexFormation}),
 the annihilation can be completed between the junctions.
 However, if the direction of the velocities are different between the two (P and P' in Fig.~\ref{fig:VortexFormation}),
 a droplet is left between the junction in the 2D diagram and then the phase $\arg\Psi_1$ winds once around the droplet,
 making a coreless vortex whose core is filled with the $\Psi_2$ component.

\begin{figure} [tpbh] \centering
  \includegraphics[width=.8 \linewidth]{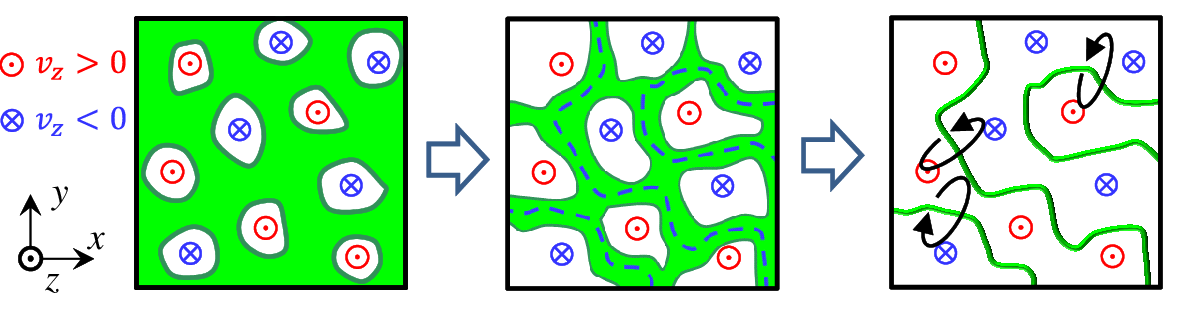}
  \caption{3D diagrams of vortex formations for $\Delta\Theta=\pi$ viewed from the positive $z$-axis.
}
\label{fig:3Dmeshed}
\end{figure}

The vortex formation strongly depends on how the junctions grow from initial fluctuations.
For example, in the experiment with small Bose-condensed clouds trapped in a spherical harmonic potential \cite{Anderson:2001},
a vortex ring emerges, affected strongly by the geometry of the system.
Here, we consider a simplest case of domain wall annihilations from initial random fluctuations in an uniform system without external potential.
 Although the growth rates of junctions with $v_z >0$ and $v_z <0$ would be different depending on $\Delta\Theta$,
 they must be statistically equivalent for $\Delta\Theta = \pi$ because of the symmetry in the configuration space of $\Psi_1$.
 Especially, if the instability is sufficiently strong,
 the fluctuations would grow monotonically and
 the vortex formation would be determined by the distribution of initial fluctuations.
 Then, a random initial fluctuation develops in a random fashion into the meshed structures as shown in Fig.~\ref{fig:3Dmeshed}.
 Since vortices emerge along the boundary between the two opposite junctions,
 the annihilation causes serpentine curves of coreless vortices.

\section{Domain wall annihilation viewed in a projected-2D space}\label{sec:4}
 The nucleated vortices stay for a while lying around the $z=0$ plane
 since the $z$ component of the velocity, induced by the random distribution of vortices, cancels out on average.
In this case,
 the annihilation process is interpreted as phase ordering dynamics of a real scalar field in two dimensions
 by {\it projecting} the three-dimensional dynamics onto the $x$-$y$ plane \cite{Takeuchi:2012}.
 When the phenomenon is viewed in the projected-2D space,
 the brane annihilation in the original 3D space may be reduced to a ferromagnetic phase transition by a rapid quench.
 We shall explain the correspondence between the 3D and projected-2D systems qualitatively in the following,
 which is summarized in the table \ref{table:3D-2D}.
 
\begin{table}[htbp] \centering
\begin{tabular}{c|c}
\hline
3D & Projected-2D  \\ \hline 
Coreless vortices & Kinks \\
Inter-brane distance $R$ & `Temperature' \\
Phase difference $\Delta\Theta \neq \pi$ & `External field' \\
 \hline
\end{tabular}
\caption{Correspondence between the original 3D system and the projected-2D system.}
\label{table:3D-2D}
\end{table}

\begin{figure} [tpbh] \centering
  \includegraphics[width=.8 \linewidth]{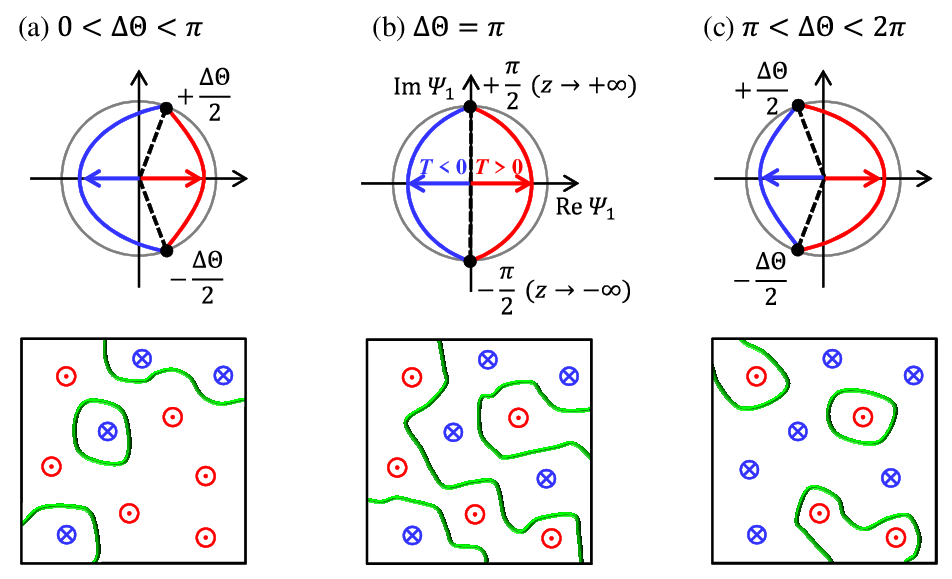}
  \caption{Schematic explanation of the $\Delta\Theta$-dependence of tachyon condensation.
  (Top) Dynamics of tachyon field due to the growth of junctions.
 Arrows represents the growth of tachyon field from the initial state (broken lines) in the configuration space of $\Psi_1$.
 The left and right solid curves show the mapping of the profiles of  the $\Psi_1$ component along junctions with $v_z<0$ ($T<0$) and $v_z>0$ ($T>0$), respectively.
 Here, we neglect the correlation between the two $\Psi_1$ domains and assume that the $\Psi_1$ component vanishes between the branes in the initial state with $T=0$.
The domain structure is statistically symmetric between $T>0$ and $T<0$ domains for (b) $\Delta\Theta=\pi$,
 but asymmetric for (a) $0\leq \Delta\Theta<\pi$ and (c) $\pi< \Delta\Theta\leq 2\pi$. 
}
\label{fig:DeltaPhase}
\end{figure}

 Remember that the inhomogeneous annihilation occurs by the growth of junctions with a superfluid current velocity $v_z>0$ or $v_z<0$.
 In this correspondence, we think that the junctions with $v_z>0$ ($v_z<0$) is induced effectively by a growth of a certain real field $T>0$ ($T<0$) in the projected-2D space.
 Here, the amplitude and the sign of the field $T$ respectively corresponds to the `amplitude' and the `direction' of the variation of $\Psi_1$'s profile from that in the initial state in the configuration space as is illustrated schematically in Fig.~\ref{fig:VortexFormation} (see also Fig.~\ref{fig:DeltaPhase}).
 The state of a pair of branes ($T=0$) represents the local maximum of the effective potential $V(T)$.
 For the symmetric case with $\Delta\Theta=\pi$, the potential should be symmetric, $V(-T)=V(T)$,
 and the growth process corresponds to tachyon condensation accompanied by spontaneous symmetry breaking, which causes `magnetization' $T>0$ or $T<0$.

 The inhomogeneous annihilation sketched in Fig.~\ref{fig:3Dmeshed} is projected as an inhomogeneous `magnetization' onto the projected-2D space. 
 Since a coreless vortex is nucleated between junctions with $v_z>0$ and $v_z<0$,
 a vortex is {\it projected} onto the 2D space as a kink,
 where the tachyon field $T$ changes its sign across the {\it shadow} of the vortex line.
 The tachyon field $T$ is constant, $T=\pm T_b$, far from the kink core,
 where the potential $V(T)$ takes the minimum value. 
 Therefore,
 the domain structure of $T$ on a length scale larger than the kink thickness in the projected-2D space is independent of how to define the tachyon field $T$, 
 although the details of the internal structure of the kinks depend on the definition of $T$.
 It was shown numerically that the time development of the domain structure of $T$ obeys actually the dynamic scaling law of phase ordering kinetics \cite{Takeuchi:2012}.

The curvature of the potential $V(T)$ at $T=0$ is related to the strength of the instability, which causes the `magnetization'.
Since the annihilation occurs due to the interaction between branes,
the instability must become stronger as the inter-brane distance $R$ decreases.
The curvature of $V(T)$ around $T=0$ becomes smaller for weaker instability for larger $R$,
 and one obtains $dV/dT=d^2V/dT^2=0$ when the inter-brane interaction vanishes completely for $R\to \infty$. 
 In terms of the Ginzburg--Landau theory for ferromagnetic phase transitions,
 $F_2\equiv d^2V/dT^2$ is a function of temperature and becomes negative below a transition temperature.
 In this sense, the inter-brane distance $R$ works as `temperature'.
Since the strength of the inter-brane interaction becomes small exponentially with $R$ and the instability should vanish exactly for $R\to \infty$,  
 the infinity distance may correspond to the transition `temperature'.

\begin{figure} [tpbh] \centering
  \includegraphics[width=.8 \linewidth]{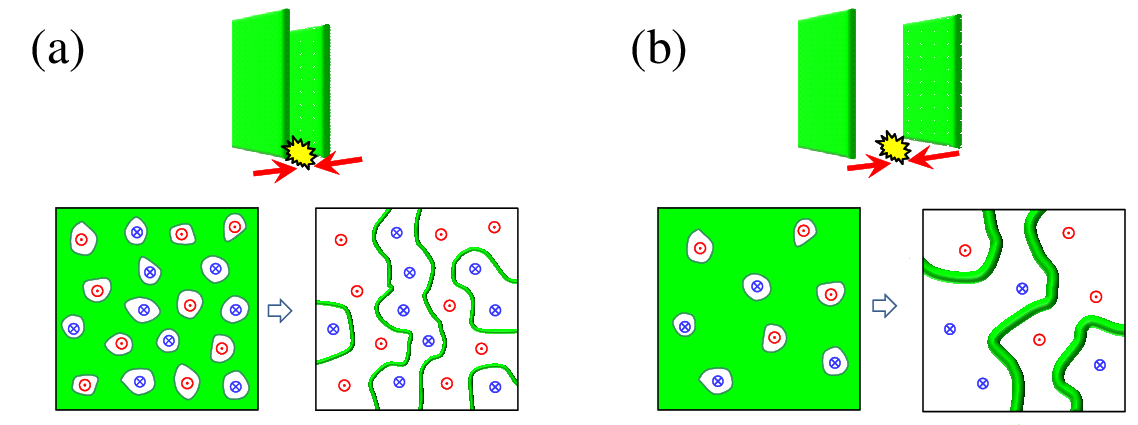}
  \caption{Schematic explanation of the $R$-dependence of tachyon condensation.
  (a) and (b) show the cases for smaller and larger $R$, respectively.
  The line density of nucleated vortices increase with decrease of the initial inter-brane distance $R$.
  The nucleated vortices becomes thicker as $R$ increases. 
}
\label{fig:R_depend}
\end{figure}

 The strength of the instability is related to the line density of vortices nucleated immediately after the collision of branes.  
 The line density can be estimated  from the wave length of the most unstable modes,
 when the instability is so strong that the growth of the mode dominates the vortex formations.
 According to Ref.~\cite{Takeuchi:2011}, the wave length of the most unstable mode, which has the maximum imaginary frequency, increases with $R$, and thus the line density  increases with decrease of $R$.
 Since the norm $N_2$ is an increasing function of $R$,
the nucleated vortices becomes thicker as $R$ increases.   
 The $R$-dependence of the vortex formation is schematically illustrated in Fig.~\ref{fig:R_depend}.

 The potential $V(T)$ becomes asymmetric, $V(-T)\neq V(T)$, for $\Delta\Theta \neq \pi$ because the amplitudes of the superfluid current velocities (\ref{eq:v_z}) induced by the junctions are different between the two velocities $v_z>0$ and $v_z<0$.
 Similarly to the analogy above, the phase difference $\Delta\Theta \neq \pi$ plays the role of the `external field' in the projected-2D system.
 Since the kinetic energy density induced by the superfluid current is roughly estimated as $\sim \frac{m}{2}\frac{\mu_1}{g}v_z^2$,
 'magnetization density' $T>0$ ($T<0$) prefers to be amplified energetically for $0 \leq \Delta \Theta<\pi$ ($\pi<\Delta \Theta \leq 2\pi$).
 Then, the domains of $T>0$ ($T<0$) tend to {\it percolate} across the projected-2D system and some domains of $T<0$ ($T>0$) forms droplets.
 The droplet corresponds to a vortex ring in the original three-dimensional space.
 Since a kink across the system is realized when both domains ($T>0$ and $T<0$) {\it percolate},
 the possibility to form vortex lines across the system is maximized for $\Delta\Theta=\pi$.
 The $\Delta\Theta$-dependence of the vortex formation is schematically illustrated in Fig.~\ref{fig:DeltaPhase}.

\section{Annihilation with bridged vortices between branes}\label{sec:5}
\begin{figure} [tpbh] \centering
  \includegraphics[width=.7 \linewidth]{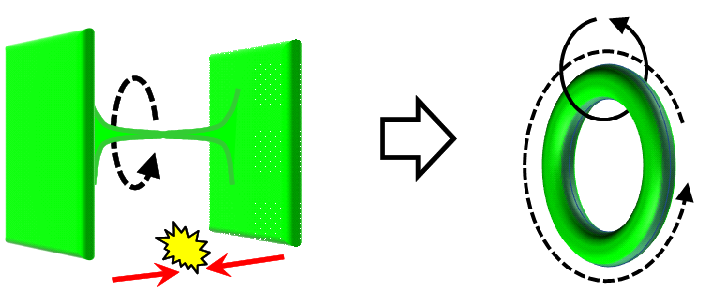}
  \caption{Schematic diagram of formation of a vorton from brane annihilation with a bridged vortex.
  Solid and broken arrows represent the directions of superfluid currents of the $\Psi_1$ and $\Psi_2$ component, respectively.
}
\label{fig:VortonBridge}
\end{figure}

 In the discussion above, we did not take into account explicitly the influence of the superfluid current due to the phase gradient of  the $\Psi_2$ component.
 The influence can be important when the population of the $\Psi_2$ component, localized along the coreless vortices, becomes larger.
 Such a situation is realized in the annihilation starting from a larger inter-brane distance,
 where vortices with flat cores are nucleated.
 Especially, when the vortices are bridged between the domain walls,
 the annihilation dynamics cause an additional nontrivial effect,
 nucleation of vortons \cite{Nitta:2012} (see Fig.~\ref{fig:VortonBridge}).
 
 When a vortex ends on the domain wall,
 the vortex exerts a force on the end point joining the brane, causing the wall or the $|\Psi_1|=|\Psi_2|$ plane to bend.
 Such a configuration is discussed by considering the analogy with string theory \cite{Kasamatsu:2010},
 where the domain wall and the vortex attached to the wall correspond to a D-brane and a fundamental string.
 If a vortex is bridged between a domain wall and an anti-domain wall,
 the bridged vortex draw both walls [see Fig.~\ref{fig:VortonBridge} (left))].

\begin{figure} [tpbh] \centering
  \includegraphics[width=.75 \linewidth]{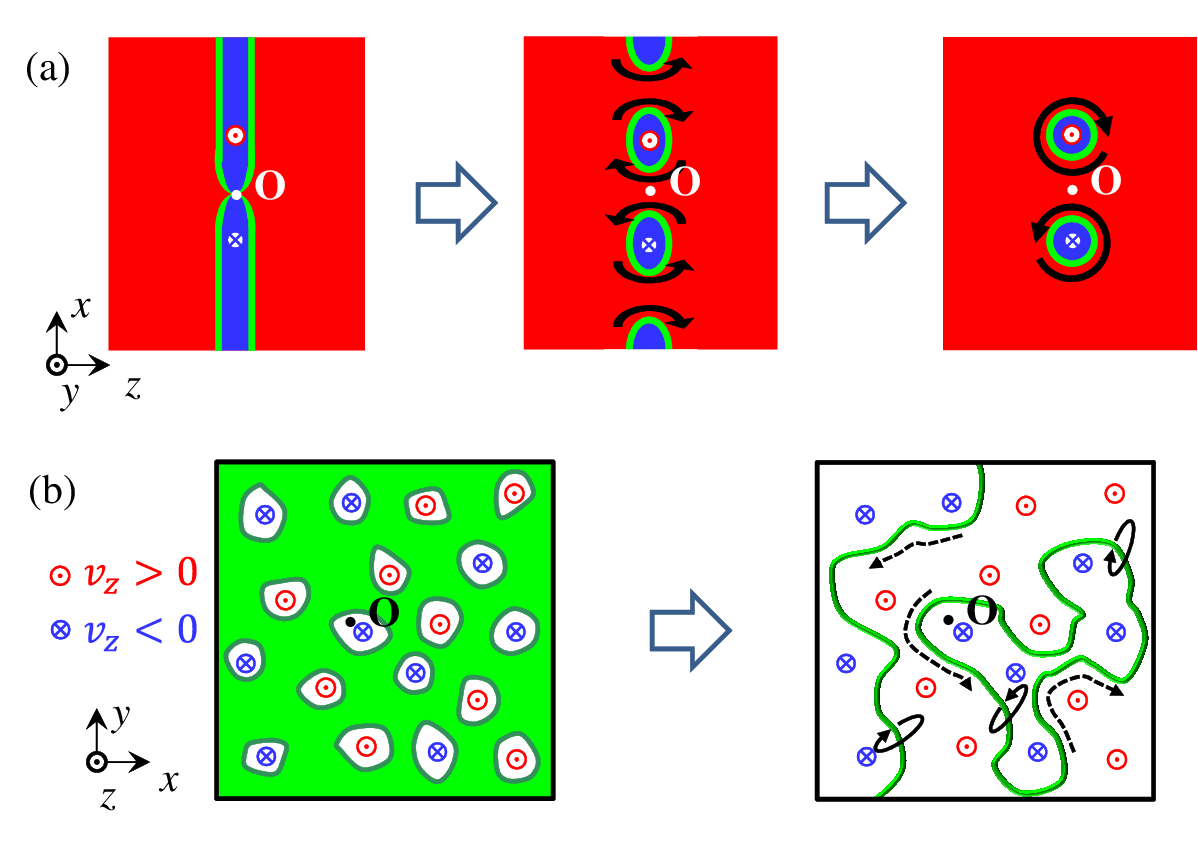}
  \caption{ (a) 2D diagrams of brane annihilation with a bridged vortex.
 A quantized vortices passes through the origin along the $z$-axis.
 The marks $\odot$ and $\otimes$ shows the directions,
 positive and negative directions of the $y$ axis, of the superfluid currents of the $\Psi_2$ component.
  Superfluid currents of the $\Psi_1$ component are represented with solid arrows. 
  (b) 3D diagrams of the brane annihilation viewed from the positive $z$-axis.
 Solid and broken arrows represent the directions of superfluid currents of the $\Psi_1$ and $\Psi_2$ component, respectively.
}
\label{fig:Vorton}
\end{figure}

 Let us consider a bridged vortex in the $\Psi_2$ component between the two $\Psi_1$ domains with a phase difference $\Delta \Theta=\pi$.
 The $\Psi_1$ component has a nodal plane, on which we have $\Psi_1=0$,
 and the bridged vortex is perpendicular to the walls passing through the nodal plane at the origin O.
 This configuration is shown schematically in the 2D diagram of Fig.~\ref{fig:Vorton}(a) (left). 
 If a junction grows along the bridged vortex [Fig.~\ref{fig:Vorton}(a) (center)],
 two coreless vortices are nucleated below and above the origin.
 Since the $\Psi_2$ component is rotating around the $z$ axis in the initial state,
 the $\Psi_2$ currents inside the vortex cores flow in opposite directions, positive direction and negative direction of the $y$ axis in the 2D diagram of Fig.~\ref{fig:Vorton}(a) (right).

In 3D diagram of Fig.~\ref{fig:Vorton}(b), an inhomogeneous growth of junctions can cause a vortex loop surrounding the origin.
Here, the $\Psi_2$ component inside the vortex core has a nontrivial winding along the loop.
Such a ``twisted'' vortex loop is called a vorton \cite{Davis:1988, Radu:2008}.
 In the case of usual vortex loop, in the presence of energy dissipation,
 the loop shrinks easily to smaller loops or rings, and the rings decay into phonons,
 because the energy of a vortex ring is a monotonically increasing function of the radius of the ring.
 However, in the case of  a vorton, a ``twisted'' vortex ring,
 the velocity of the $\Psi_2$ current along the vortex ring is inversely proportional to the ring radius
 if the winding number of the $\Psi_2$ component along the ring is fixed.
Thus, the vorton can be stabilized with a certain radius to reduce the kinetic energy due to the $\Psi_2$ current.

The `twisted' loops can be unstable due to hydrodynamics instability in the presence of a relative velocity between the two components.
 The most likely candidate is the Kelvin-Helmholtz instability,
 which causes nucleation of vortices from the interface between the two phase-separated condensates \cite{Takeuchi:2010}.
 The instability may causes vortices in the $\Psi_2$ component leading to a reduction of the winding number of the `twisted' loops. 
 For simplicity, we neglect such an effect and consider dynamics of the vorton loops in the projected-2D space by assuming that their winding number is conserved during the dynamics.

 Here, we discuss reconnection of the kink loops in the 2D system.
 A twisted loop, along which the $\Psi_2$ current is winding $n$ times, is called an $n$-loop, where $n$ is an integer.
 An $n$-loop with $n>0$ ($n<0$) is realized, for example, when $|n|$ bridged vortices with a circulation quantum number $+1$ ($-1$) are surrounded by the loop.
 Here, we do not consider a kink lying across the system
 because the $\Psi_2$ component flows along the kink out to the edge of the projected-2D system,
 the surface of the atomic cloud in the original 3D system, and then the winding number of the kink is not well defined.

\begin{figure} [tpbh] \centering
  \includegraphics[width=.8 \linewidth]{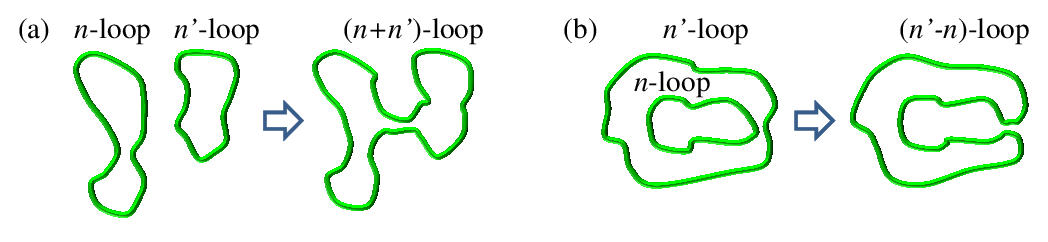}
  \caption{Reconnection of twisted loops in the projected-2D space.
 The integers $n$, $n'$, $n+n'$, and $n'-n$ represent the winding number of the $\Psi_2$ current along each loops.
 (a) A reconnection between an $n$-loop and an $n'$-loop results in an $(n+n')$-loop.
 (b) When an $n'$-loop surrounds an $n$-loop, a reconnection between them leads to an $(n'-n)$-loop. 
}
\label{fig:Reconnection}
\end{figure}

 A reconnection between twisted loops is classified typically into two cases by considering the winding number.
 The first case is that a reconnection between the two loops, either of which does not surround the other loop [Fig.~\ref{fig:Reconnection}(a)].
 In this case, a reconnection between an $n$-loop and an $n'$-loop makes an $(n+n')$-loop.
 In the second case where an $n'$-loop surrounds an $n$-loop,
 the reconnection between the loops results in an $(n'-n)$-loop [Fig.~\ref{fig:Reconnection}(b)].
 Such nontrivial effects can change the statistical properties of the phase ordering dynamics in the projected-2D system.
 For example, the statistic probability of reconnections between loops may depends on their winding numbers,
 which would influence the dynamic scaling law or the decay law of kink density in phase ordering kinetics \cite{Bray:1994}.

 \section{Conclusions and discussion}\label{sec:6}
Vortex formations via the pair annihilation of domain walls in two-component BECs are regarded as kink formations in the effective tachyon field theory of the projected-2D system. The tachyon potential is characterized by the initial inter-brane distance and the phase difference between the two bulks separated by the brane pair. In the projected-2D system, the inter-brane distance and the phase difference play the roles of the temperature and external field in ferromagnetic systems, respectively.
 Vortons can be nucleated via an annihilation of an brane and an anti-brane with bridged vortices.
 
 These phenomena can be monitored directly in experiments of cold atom systems with the techniques used in Ref.~\cite{Anderson:2001}.
 The formation of serpentine vortices and the effective phase ordering dynamics are feasible by using a trapped condensate whose size along the branes is much larger than the characteristic wavelength $\sim\xi$ of the tachyon mode.
The domain structure of the tachyon field can be directly visualized by observing the density depletion in the $\Psi_1$ component due to vortices in the central slice of the atomic cloud \cite{Andrew:1997}.
It is also interesting that a similar situations is experimentally realized in superfluid $^3$He \cite{Bradley:2008a,Bradley:2008b}.
 It will be fruitful to reconsider the experiments from the viewpoint discussed here. 
We hope that future experiments reveal, for example, how quantum and thermal fluctuations influence the defect formations and the relaxation dynamics in the projected-2D system beyond the mean field theory,
which must be useful to establish the SSB phenomena occurring in a restricted lower-dimensional subspace.

\begin{acknowledgements}
This work was supported by KAKENHI from JSPS 
(Grant Nos. 21340104, 21740267 and 23740198).
This work was also supported
by the ``Topological Quantum Phenomena'' 
(Nos. 22103003 and 23103515)
Grant-in Aid for Scientific Research on Innovative Areas 
from the Ministry of Education, Culture, Sports, Science and Technology 
(MEXT) of Japan. 
\end{acknowledgements}

\end{document}